\documentclass[aps, prd, twocolumn, amsmath, amssymb, nofootinbib, floatfix, superscriptaddress]{revtex4-1}

\usepackage[english]{babel}
\usepackage{color}
\usepackage{comment}
\usepackage{graphicx}
\usepackage{placeins}  
\usepackage[normalem]{ulem}

\newcommand{\eq}{\begin{equation}}
\newcommand{\en}{\end{equation}}
\newcommand{\eqa}{\begin{eqnarray}}
\newcommand{\ena}{\end{eqnarray}}
\newcommand{\Scl}{S_{\mbox{\tiny{cl}}}}
\newcommand{\tmb}[1]{{\mbox{\tiny{#1}}}}
\def\sumn{\sum_{n=1}^{\infty}}

\begin{document}

\title{The unreasonable effectiveness of effective string theory:\\ the case of the 3d SU(2) Higgs model}

\author{Claudio~Bonati}
\email{claudio.bonati@unipi.it}
\affiliation{Dipartimento di Fisica dell'Universit\`a di Pisa and INFN, Pisa \\ Largo Pontecorvo 3, I-56127 Pisa, Italy.}

\author{Michele~Caselle}
\email{caselle@to.infn.it}
\affiliation{Dipartimento di Fisica dell'Universit\`a di Torino and INFN, Turin \\ Via Pietro Giuria 1, I-10125 Turin, Italy.}

\author{Silvia~Morlacchi}
\email{silvia.morlacchi@sns.it}
\affiliation{Scuola Normale Superiore \\ Piazza dei Cavalieri 7, I-56126 Pisa, Italy}

\date{\today}

\begin{abstract}
We study string breaking in the three dimensional SU(2) Higgs model, using
values of the gauge coupling for which the confinement-like and Higgs-like
regions of the phase diagram are separated just by a smooth crossover.
We show that even in the presence of string breaking, the confining
part of the interquark potential is well described by the Effective String
Theory  and that also the fine details of the effective string, like the higher
order terms of the Nambu-Goto action or the boundary correction, can be
precisely extracted from the fits and agree with the effective string
predictions. We comment on the implications of these results
for QCD simulations with dynamical quarks.
\end{abstract}

\maketitle

\section{Introduction}
\label{sec:introduction}

A powerful tool to describe the nonperturbative behaviour of the interquark
potential in confining gauge theories is the so called ``Effective String
Theory''(EST) in which the confining flux tube joining together a
static quark-antiquark pair is modeled as a thin vibrating
string~\cite{Nambu:1974zg, Goto:1971ce, Luscher:1980ac, Luscher:1980fr,
Polchinski:1991ax}.  This approach has a long history (for a review see for
instance \cite{Aharony:2013ipa, Brandt:2016xsp, Caselle:2021eir}) and has been
shown to be a highly predictive effective model, whose results can be
successfully compared with the most precise existing Montecarlo simulations in
Lattice Gauge Theories (LGTs).  

There are two main reasons for the great phenomenological success of this
approach. The first is that EST is strongly constrained by the Lorentz symmetry
and is thus much more predictive that typical effective theories. The second is
that the range of validity of EST is precisely defined (see below for a
detailed derivation) and is thus possible to compare EST predictions with
numerical data in a controlled and unambiguous way.

The EST description of long flux tubes is perfectly natural to study
long distance properties of pure gauge theories, however a major issue in this
context is to understand if the EST approach can be extended also beyond pure
gauge theories. In view of a possible application to QCD, it would be important
to understand which is the fate of the EST description in the presence
of dynamical matter fields and thus in a string breaking
scenario.

A perfect laboratory to address this issue is the $SU(2)$ Higgs model in three
dimensions, which is very similar to real QCD for what concerns string
breaking, but at the same time can be simulated at high precision with
relatively small effort. The main goal of this paper is to explore the
confining potential of the model in the crossover region of the phase diagram
between the confining regime and the broken string regime and compare the
results of the simulations with the EST predictions.  This paper is the natural
continuation of the analysis initiated in \cite{Bonati:2020orj} where,  in a
model similar to the one discussed here, the shape and size of the confining
flux tube was compared with EST predictions.

As we shall see, the confining part of the potential is perfectly described by
EST even in its fine details. In particular the contribution due to the higher
order terms beyond the Gaussian one contained in the Nambu-Goto action and the
so called ``boundary term'' of the EST are in perfect agreement with the data.
Moreover we show that if one tries to fit the data neglecting the information
coming from the EST action, a wrong value for the string tension  (which plays
a central role in modelling quarkonia spectra) is obtained. We guess
that a similar scenario should occur also in real QCD \cite{Bali:2005fu,
Bulava:2019iut} and, in view of the recent efforts to model quarkonia spectra
using high precision lattice results \cite{Bicudo:2019ymo, Bicudo:2020qhp,
Bruschini:2020voj, Bruschini:2021sjh}, we stress the importance of the
inclusion of EST corrections in the potential models. The role of these
corrections will become more and more important as the precision of QCD
simulations with dynamical quarks will improve and it will be mandatory to keep
them into account for a reliable description of mesonic states in the confining
regime of QCD.

This paper is organized as follows: we devote Sect.~2 to a description of the
model and Sect.~3 to a brief summary of EST results. Our main results are
collected in Sect.~4, while Sect.~5 is devoted to a few concluding remarks.

\section{The model}

To study the three dimensional non abelian Higgs model, with gauge group
$\mathrm{SU}(N_c)$ and $N_f$ scalar fields transforming in the fundamental
representation of the gauge group, we can use the following discretization
\begin{equation}\label{action}
\begin{aligned}
S  = &- N_f \beta_h\sum_{x,\mu} {\rm Re}\, {\rm Tr}
\left(\varphi_{x}^\dag  U_{x,\mu} \varphi_{x+\hat{\mu}}\right) \\
&- \frac{\beta}{N_c}  \sum_{x, \mu>\nu} {\rm Re} {\rm Tr}\Pi_{\mu\nu}(x) \ .
\end{aligned}
\end{equation}
Here ${x}$ denotes a point of a three dimensional isotropic
lattice with periodic boundary conditions, $\mu,\nu\in \{0, 1, 2\}$ label the
lattice directions, and  ${\rm Re} {\rm Tr}\Pi_{\mu\nu}(x)$, with 
\begin{equation}\label{eq:plaq}
\Pi_{\mu\nu}(x)=\,\big[U_{\mu}(x)U_{\nu}(x+\hat{\mu})
U^{\dag}_{\mu}(x+\hat{\nu})U^{\dag}_{\nu}(x)\big]
\end{equation}
denotes the standard Wilson action, i.e.  the trace of the product of the link
variables around the plaquette in position $x$ laying in the plane $(\mu,
\nu)$. In eq.(\ref{action})  $U_{x, \mu}$ is a matrix belonging to
the $\mathrm{SU}(N_c)$ group while  the $\varphi_{x}$ fields are
$N_c\times N_f$ complex matrices which satisfy the constraint $\mathrm{Tr}
~\varphi_{x}^{\dag}\varphi_{x}=1$. 

String breaking is present in this model for any positive value of $N_f$ as
soon as\footnote{Note that gauge field correlators are symmetric for
$\beta_h\to-\beta_h$, as follows from the change of variable $\varphi_{x}\to
(-1)^{x_1+x_2+x_3}\varphi_{x}$. For this reason we consider just positive
values of $\beta_h$ in the following.} $\beta_h> 0$, and in the following we
consider the case $N_f=1$. There are two reasons for this choice: first of all
the $N_f=1$ case is the simplest one from the computational point of view.
Moreover in this case it can be rigorously shown that a single thermodynamic
phase exists \cite{Osterwalder:1977pc, Fradkin:1978dv}, while for $N_f>1$ a
global $SU(N_f)$ symmetry is present, which gets spontaneously broken for large
enough values of $\beta_h$ \cite{Bonati:2019zrt, Bonati:2020elf}.  The presence
of the phase transition between the $\mathrm{SU}(N_f)$ disordered/ordered
phases for $N_f>1$ introduces additional features beyond string breaking, which
could hinder the possibility of making contact with real world QCD.  

For the sake of the simplicity we study the $N_c=2$ case, which is the
computationally easiest model of this class.  For $N_f=1$ and $N_c=2$ (and on
an infinite lattice) the model in eq.~(\ref{action}) reduces, in the limit
$\beta\to\infty$, to the standard discretization of the nonlinear O(4)
$\sigma$-model, for which a second order phase transition is known to exist
at $\beta_h=0.93586(8)$ \cite{Campostrini:1995np, Ballesteros:1996bd}. A
priori, for large values of $\beta$ a line of first order phase transition
could be present, which however must end for some $\beta_c>0$, like in the
four-dimensional version of the model (see e.g.  \cite{Bonati:2009pf}).  For
this reason in the following we explicitly check the absence of phase
transitions in the region of the parameter space investigated. 

We are interested in the interquark potential $V(R)$ which can be extracted
from the correlator of Polyakov loops
\eq
\langle P(x)P^\dagger(x+R) \rangle ~\equiv ~{\rm e}^{- N_t V(R)}\ ,
\label{polya}
\en
\noindent
as follows:
\eq
V(R)=-\frac{1}{N_t}\log{\langle P(x)P^\dagger(x+R) \rangle  } \ .
\label{potft}
\en
where we set the lattice spacing $a$ to $1$ and it will be implied in the
following, $N_t$ denotes the lattice size in the compactified time direction
and we are studying the system on a cubic lattice with the same size in the
spacelike directions (which we shall denote as $L$) and in the time direction.

The pure gauge limit ($\beta_h=0$) of the model we consider here has been the
subject of several studies in the past \cite{Ambjorn:1984me, Teper:1998te,
Caselle:2004er, Caselle:2011vk, Bringoltz:2006zg, Brandt:2010bw,
Brandt:2017yzw, Brandt:2018fft, Brandt:2021kvt} since it is the simplest LGT
with a non-abelian continuous gauge group and is thus a perfect laboratory to
test large distance, non perturbative, features of these theories.  

Similarly, the model in the presence of an external bosonic field
($\beta_h > 0$) has been used a lot in the past to understand and model string
breaking.  Indeed the choice of bosonic instead of fermionic fields represents
an enormous simplification from the numerical point of view, while keeping
essentially unchanged the phenomenology of string breaking. Thus one may hope
to use models like the one  we discuss in this paper as toy models to better
understand the string breaking phenomenon in real QCD.

In order to make contact with previous studies in this context
\cite{Philipsen:1996af, Philipsen:1997rq, Philipsen:1998de, Knechtli:1998gf,
Knechtli:2000df}, let us stress that, while in those studies the external field
had a self-interaction term of the type $\lambda \phi^4$, in our model the
self-interaction term is substituted by the $\mathrm{Tr}
~\varphi_{\textbf{x}}^{\dag}\varphi_{\textbf{x}}=1$ constraint. In this respect
we may consider our model as the $\lambda\to \infty$ limit of those studied in
the past. This does not change the phenomenology of string breaking but has the
advantage of eliminating the $\lambda$ parameter from the game.

A different strategy to investigate string breaking would be to study the
potential, or the flux tube, between static charges in higher representations
of the gauge group \cite{Stephenson:1999kh, Philipsen:1999wf,
Kratochvila:2003zj, Kallio:2000jc, Pepe:2009in, Bonati:2020orj}. A disadvantage
of this kind of approach is however that there is no way of tuning the physical
distance at which string breaking happens, and it could be that EST never
apply in this case (the precise range of validity of EST will be reviewed in
the next section).  Using dynamical matter fields in the fundamental
representation we can instead vary the physical string breaking length by
changing the coupling between matter and gauge fields, i.e. $\beta_h$ in
eq.~(\ref{action}).

One of the advantages of studying the $SU(2)$ model in (2+1) dimensions is that
we can leverage on previous studies to fix the parameters of the model.  In
particular we can use the scale setting expression obtained in
\cite{Teper:1998te}   
\begin{equation}\label{scale_setting}
a\sqrt{\sigma}=\frac{1.324(12)}{\beta}+\frac{1.20(11)}{\beta^2}+\mathcal{O}(\beta^{-3})\ ,
\end{equation}
which is expected to be valid for $\beta\ge 4.5$.  Moreover we shall fix in the
following $\beta=9.0$ for which a high precision study of the interquark
potential can be found in \cite{Caselle:2004er}.

\begin{figure}[t] 
\includegraphics[width=0.85\columnwidth, clip]{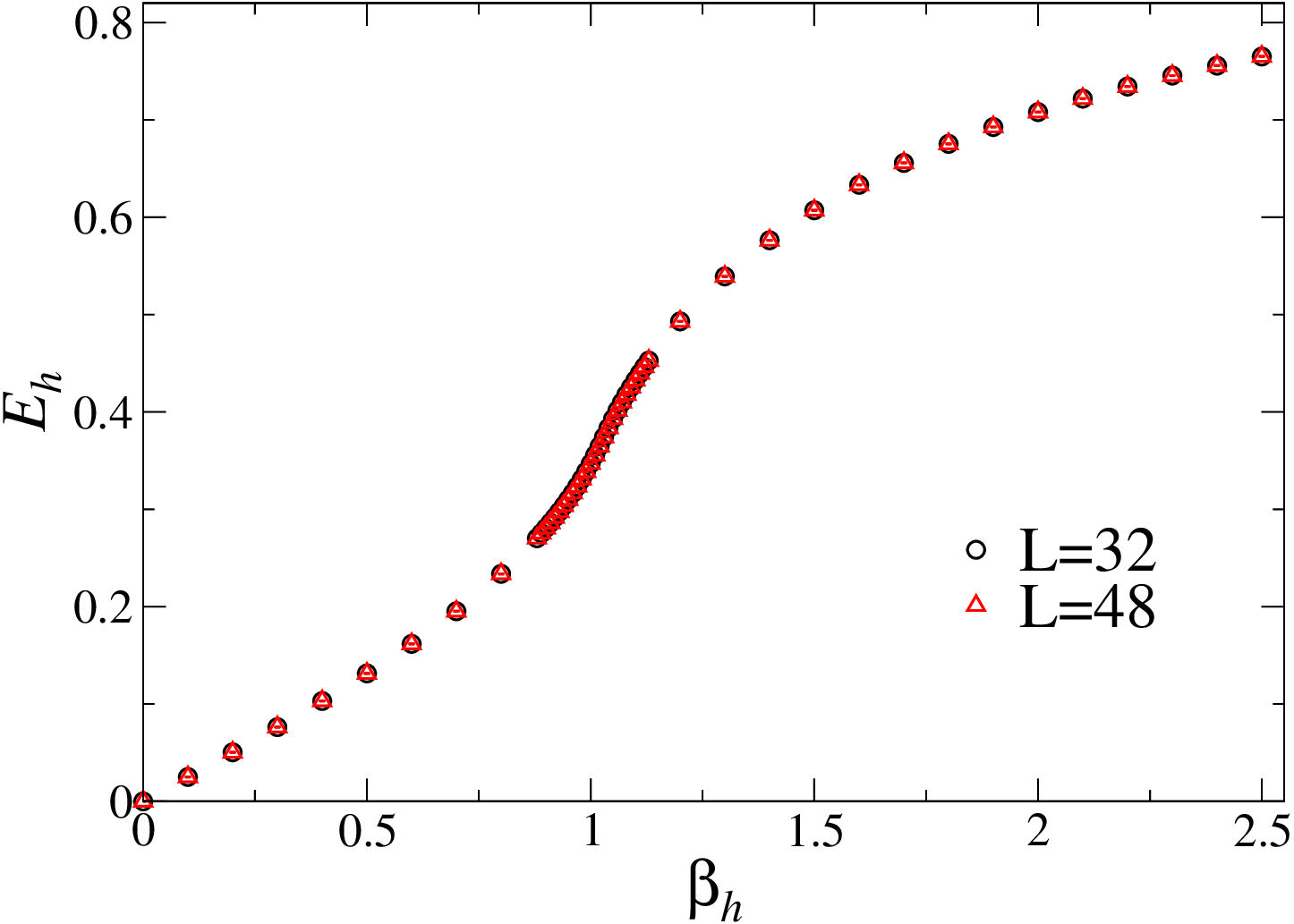}
\includegraphics[width=0.85\columnwidth, clip]{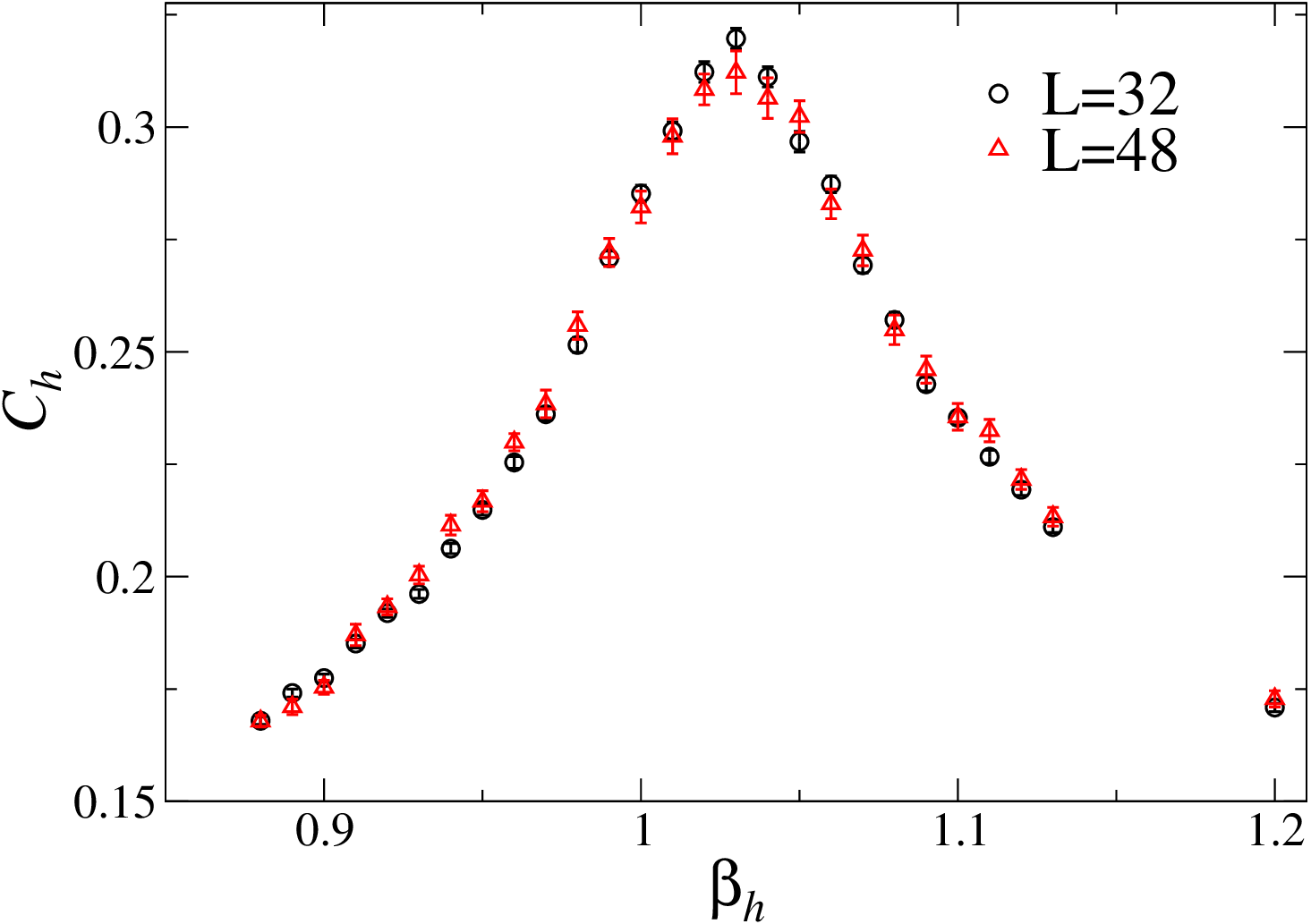}
\caption{Dependence of $E_h$ and $C_h$ (defined in eq.~\eqref{eh_chih}) on
$\beta_h$ for $\beta=9$. Two different lattice sizes are shown, $L=32$ and
$L=48$, and no signal of phase transition is present.}
\label{fig:betah_scan}
\end{figure}

As a first step of our analysis we verified that for our choice of
$\beta,\beta_h$ no phase transition is encountered. To this
end we performed for $\beta=9.0$ a scan in $\beta_h$,
monitoring the observables
\begin{equation}\label{eh_chih}
\begin{aligned}
&E_h=\langle \mathrm{Re}\big(\varphi_{x}^\dag  U_{x,\mu} \varphi_{x+\hat{\mu}}\big) \rangle\ \\
&C_h= L^3 \Big(\langle \mathrm{Re}\big(\varphi_{x}^\dag  U_{x,\mu} \varphi_{x+\hat{\mu}}\big)^2 \rangle
-E_h^2 \Big)\ .
\end{aligned}
\end{equation}
This preliminary test showed the absence of phase transitions, as can be seen in
Fig.~\ref{fig:betah_scan}, where the values of $E_h$ and $C_h$ are displayed
for $L=32$ and $L=48$. A peak in the susceptibility $C_h$ is present for
$\beta_h\approx 1.04$, which however does not grow/shrinks when increasing the
lattice size, and just signals the crossover from the confinement-like to the
Higgs-like regions of the phase diagram. 

Then, for a selection of values of $\beta_h$ across the bump shown in
Fig.~\ref{fig:betah_scan}, we evaluated the Polyakov loop correlators in the
range $1\leq R\leq 20$ using a $42^3$ lattice, and from that the interquark
potential $V(R)$.  Poyakov loop correlators have been estimated using the
multihit \cite{Parisi:1983hm} and multilevel \cite{Luscher:2001up} error reduction
techniques. In all the cases the optimal number of hits was around 10 and the
optimal configuration for the multilevel was a single level scheme, with
temporal slices of 6 lattice spacings. The optimal number of updates to be
performed in the multilevel displayed instead some significant dependence on
the value of $\beta_f$ (and obviously on $R$), going from 16000 for $R>12$ at
$\beta_h=0$ to 1000 for $\beta_h=1.1$ at the same $R$. For each value of $R$
and $\beta_h$ a statistics of the order of a few thousands independent draws
was accumulated.

\begin{figure}[tb]
\includegraphics[width=0.99\columnwidth, clip]{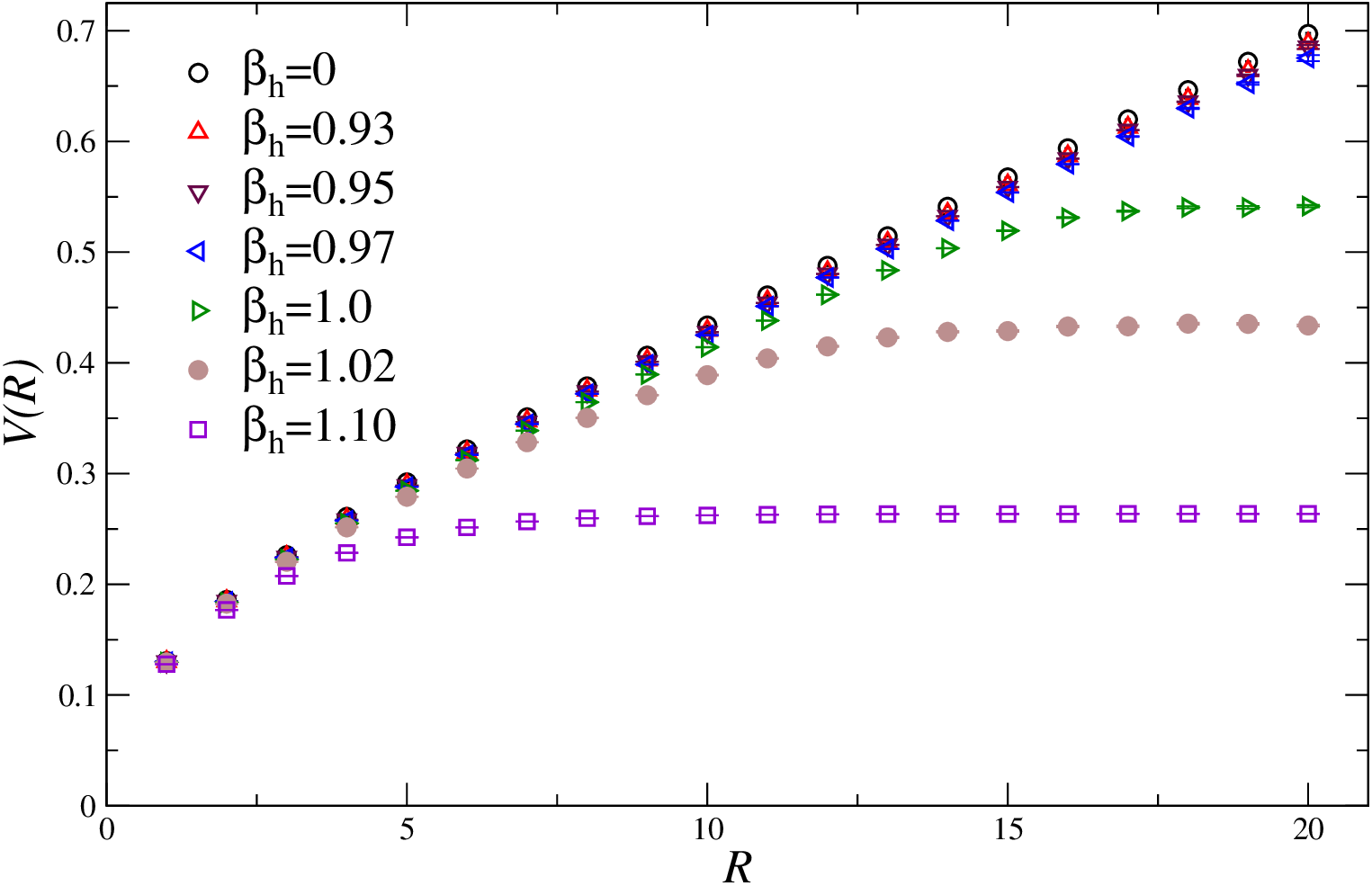}
\caption{Static potential computed on a $42^3$ lattice at $\beta=9.0$, for 
some values of $\beta_h$.}\label{fig:pot} 
\end{figure}

Results for $V(R)$ are reported in Tab.~\ref{tabV1} and are plotted in
Fig.~\ref{fig:pot}, from which we see that as $\beta_h$ increases the string
breaking phenomenon in the potential becomes more and more dramatic. In the
following sections we shall study in detail this phenomenon and will try to
model the rising part of the potential using the effective string approach. 

\begin{table*}[tbh]
\begin{center}
\begin{tabular}{llllllll}
$R$ & $\beta_h=0$   & $\beta_h=0.93$ & $\beta_h=0.95$ &  $\beta_h=0.97$  & $\beta_h=1.00$ & $\beta_h=1.02$ & $\beta_h=1.1$ \\  
1   &  0.130591(5)  & 0.130183(2)    & 0.130096(2)    &  0.129985(3)     & 0.129709(3)    & 0.129415(5)    & 0.127854(3)  \\    
2   &  0.185930(24) & 0.185026(9)    & 0.184810(8)    &  0.184482(12)    & 0.183663(10)   & 0.182627(16)   & 0.176884(8)  \\  
3   &  0.226336(54) & 0.225036(19)   & 0.224641(20)   &  0.224094(29)    & 0.222498(22)   & 0.220379(37)   & 0.207569(10) \\
4   &  0.26088(10)  & 0.258935(12)   & 0.258390(16)   &  0.257555(13)    & 0.255069(16)   & 0.251504(33)   & 0.228433(13) \\
5   &  0.292121(64) & 0.289900(19)   & 0.289224(26)   &  0.288047(20)    & 0.284551(24)   & 0.279116(56)   & 0.242479(21) \\
6   &  0.321826(94) & 0.319235(29)   & 0.318388(39)   &  0.316893(29)    & 0.312243(35)   & 0.304588(86)   & 0.251441(29) \\
7   &  0.35075(11)  & 0.347560(32)   & 0.346559(41)   &  0.344785(37)    & 0.338734(39)   & 0.32833(11)    & 0.256681(36) \\
8   &  0.378806(75) & 0.375279(42)   & 0.374005(56)   &  0.371982(48)    & 0.364514(50)   & 0.35031(17)    & 0.259691(41) \\
9   &  0.406425(98) & 0.402511(52)   & 0.400995(68)   &  0.398602(62)    & 0.389540(61)   & 0.37084(25)    & 0.261506(37) \\
10  &  0.43374(10)  & 0.429290(63)   & 0.427730(77)   &  0.424990(72)    & 0.414177(65)   & 0.38892(32)    & 0.262294(39) \\
11  &  0.46088(10)  & 0.455951(49)   & 0.454098(62)   &  0.451205(59)    & 0.438179(54)   & 0.40408(44)    & 0.262873(48) \\
12  &  0.48766(10)  & 0.482307(62)   & 0.480309(74)   &  0.477093(72)    & 0.461537(71)   & 0.41494(59)    & 0.263198(49) \\
13  &  0.51430(11)  & 0.508727(86)   & 0.506574(97)   &  0.502830(96)    & 0.48359(10)    & 0.42295(64)    & 0.263385(47) \\
14  &  0.54099(14)  & 0.53490(10)    & 0.53249(12)    &  0.52847(11)     & 0.50355(16)    & 0.42801(70)    & 0.263563(50) \\
15  &  0.56751(16)  & 0.56067(13)    & 0.55871(15)    &  0.55391(13)     & 0.51934(39)    & 0.42872(77)    & 0.263498(49) \\
16  &  0.59377(19)  & 0.58691(17)    & 0.58445(18)    &  0.57948(18)     & 0.53115(49)    & 0.43282(79)    & 0.263496(56) \\
17  &  0.61988(24)  & 0.61270(21)    & 0.61017(24)    &  0.60444(30)     & 0.53697(60)    & 0.43306(81)    & 0.263646(56) \\
18  &  0.64631(36)  & 0.63865(32)    & 0.63577(39)    &  0.62988(53)     & 0.5405(10)     & 0.43537(83)    & 0.263618(56) \\ 
19  &  0.67184(51)  & 0.66380(48)    & 0.65982(69)    &  0.6522(11)      & 0.5404(12)     & 0.43517(83)    & 0.263622(56) \\
20  &  0.69700(95)  & 0.6884(10)     & 0.6853(17)     &  0.6752(27)      & 0.5416(10)     & 0.43378(86)    & 0.263646(56) 
\end{tabular}
\end{center}
\caption{Values of the static potential (in lattice units) for $\beta=9.0$ and several $\beta_h$,
computed on a $42^3$ lattice. Note that values corresponding to different $R$s are independent from each other, 
since they have been estimated using different runs.}
\label{tabV1}
\end{table*}

\section{Effective string predictions}

Even if a rigorous proof of quark confinement in Yang-Mills theories is still
missing, there is little doubt that confinement  is associated to the formation
of a thin string-like flux tube~\cite{Nambu:1974zg, Goto:1971ce,
Luscher:1980ac, Luscher:1980fr, Polchinski:1991ax}, which generates, for large
quark separations, a linearly  rising  confining potential.

The simplest example of an EST leading to a linearly rising potential was
proposed more than forty years ago by L\"uscher and
collaborators~\cite{Luscher:1980ac, Luscher:1980fr}. They suggested to model
the fluctuations of the flux tube in the transverse directions as a free
massless bosonic field theory in two dimensions.  
\eq
S[X]=S_{cl}+S_0[X]+\dots,
\label{frees}
\en
where the classical action $S_{cl}$ describes the usual perimeter-area term,
$X$ denotes the two-dimensional bosonic fields $X_i(\xi_1,\xi_2)$, with
$i=1,2,\dots, D-2$,   where $D$ is the number of spacetime dimensions (in our
case $D=3$), $D-2$ is the number of transverse directions, $\xi_1,\xi_2$ are
the coordinates on the world-sheet, $S_0[X]$ is the Gaussian action
\eq
S_0[X]=\frac{\sigma}{2}\int d^2\xi\left(\partial_\alpha X\cdot\partial^\alpha X
\right) ~
\label{gauss}
\en
and we are assuming an Euclidean signature for both the worldsheet and the
target space.  The fields $X_i$ describe the transverse displacements of the
string with respect the configuration of minimal energy.

The gaussian action can be easily integrated, leading to an explicit expression
for the interquark potential, which in the large distance limit is 
\eq
V(R)=\sigma R  + c -\frac{\pi(D-2)}{24R}+O(1/R^2)~.
\label{freeV}
\en
where $\sigma$ denotes, as usual, the string tension and $c$ is related to the
``perimeter'' term mentioned above and keeps into account the classical
contribution of the Polyakov loops to the potential. We see from the above
equation that the effect of the string fluctuations is  a correction, known as
``L\"uscher term'', proportional to $1/r$  to the linearly rising potential.
This is the first example of an effective string action and, as we shall see
below, it is actually nothing else than the large distance limit of the
Nambu-Goto string written in the so called ``physical gauge''.

\subsection{The Nambu Goto action}

A careful inspection shows however that the free bosonic action is not
invariant under Lorentz transformations and that further higher order terms
must be added to grant invariance. The simplest EST which fulfills Lorentz
invariance is the Nambu Goto action~\cite{Nambu:1974zg,Goto:1971ce}:
\begin{align}
\label{NGaction}
S_\tmb{NG}= \sigma \int_\Sigma d^2\xi \sqrt{g}\ ,
\end{align} 
where $~g\equiv \det g_{\alpha\beta}~$ and
\begin{align}
\label{NGaction2}
g_{\alpha\beta}=\partial_\alpha X_\mu~\partial_\beta X^\mu
\end{align} 
is the induced metric on the reference world-sheet surface $\Sigma$ and, as
above, we denote the worldsheet coordinates as $\xi\equiv(\xi^0,\xi^1)$. This
term has a simple geometric interpretation: it measures the area of the surface
spanned by the string in the target space and is thus the natural EST
realization of the sum over surfaces weighted by their area in the rough phase
of the LGT. This model has only one free parameter:  the string tension
$\sigma$ and is thus, as we anticipated,  highly predictive.

It is easy to see \cite{Aharony:2013ipa, Brandt:2016xsp, Caselle:2021eir} that
the free bosonic action of eq.(\ref{gauss}) is the large distance limit of the
Nambu-Goto string written in the so called ``physical gauge''.  We report here
for completeness the first few terms of the expansion
\begin{equation}
\begin{aligned}
S=\Scl+\frac{\sigma}{2}\int d^2\xi\Big[&\partial_\alpha X_i\cdot\partial^\alpha X^i+
\frac{1}{8}(\partial_\alpha X_i \cdot\partial^\alpha X^i)^2 - \\
&-\frac{1}{4}(\partial_\alpha X_i \cdot\partial_\beta X^i)^2+\dots\Big]\ .
\end{aligned}
\label{action2}
\end{equation}
Despite its apparent complexity the Nambu-Goto action can be integrated exactly
in all the geometries which are relevant for LGT: the rectangle (Wilson loop)
\cite{Billo:2011fd}, the cylinder (Polyakov loop correlators)
\cite{Luscher:2004ib, Billo:2005iv} and the torus (dual interfaces)
\cite{Billo:2006zg} leading to a spectrum of states which, in the particular
case in which we are interested in this paper, i.e. the correlator of two
Polyakov loops is:
\begin{equation}
  {E}_n=\sigma R N_t
  \sqrt{1+\frac{2\pi}{\sigma R^2}\left[-\frac{1}{24}\left(D-2\right)+n\right]}\ .
\label{energylevels}
\end{equation}

In the large distance limit the spectrum is dominated 
by the lowest state $E_0$ from which we may extract the interquark potential
\eq
V(R)=c+ \sigma R \sqrt{1 -\frac{\pi(D-2)}{12~\sigma R^2}}~.
\label{NGV}
\en
and we see, as anticipated, that the L\"uscher term of eq.(\ref{freeV}) is nothing else than the first order term of 
the large distance expansion of the Nambu-Goto potential.
From eq.(\ref{NGV}) we may also obtain the
domain of validity of the EST approximation, which is given by the value $R_c$
for which the argument of the square root vanishes:
\begin{equation}
\label{Rc}
R_c=\sqrt{\frac{\pi(D-2)}{12~\sigma}}\ .
\end{equation}

\subsection{Beyond Nambu-Goto: the boundary correction}

The Nambu-Goto action is a useful approximation of the ``true'' EST which
describes the nonperturbative regime of LGTs, but it cannot be the exact
answer. First, it would predict exactly the same behaviour for any confining
LGT, without dependence on the gauge group. Second, it would predict a mean field
exponent for the deconfinement transition, in clear contradiction with LGT
simulations. It is thus of great theoretical interest to study the terms in the
EST action beyond the Nambu-Goto one. The requirement of Lorentz invariance
strongly constrains the set of allowed terms. It turns out that the leading
correction beyond Nambu-Goto is represented by the so called ``boundary term''.

This term is due to the presence of the Polyakov loops at the boundary of the
correlator.  The classical contribution associated to this correction is the
constant term $c$  which appears in the potential.  Beyond this classical term
we may find quantum corrections due to the interaction of the Polyakov loop
with the flux tube. 

The first boundary correction compatible with Lorentz invariance is \cite{Billo:2012da}
\eq
b_2\int d\xi_0 \left[
\frac{\partial_0\partial_1 X\cdot\partial_0\partial_1 X}{1+\partial_1 X\cdot\partial_1X}-
\frac{\left(\partial_0\partial_1 X\cdot\partial_1 X\right)^2}
{\left(1+\partial_1 X\cdot\partial_1X\right)^2}\right]\,.
\label{firstb}
\en 
with an arbitrary, non-universal coefficient $b_2$.
The lowest order term of the expansion of eq.(\ref{firstb}) is:
\begin{equation}
\label{derexpsb1}
S_{b,2}^{(1)} = b_2\int d\xi_0 
(\partial_0\partial_1 X)^2~
\end{equation}
The contribution of this term to the interquark potential was evaluated in
\cite{Aharony:2010cx} using the zeta function regularization:
\eq
\label{bound}
\langle S^{(1)}_{b,2} \rangle=-b_2\frac{\pi^3 N_t}{60 R^4} E_4(e^{-\frac{\pi N_t}{ R}})
\en
where $E_4$ denotes the fourth order Eisenstein series 
\begin{equation}\label{series1}
\begin{aligned}
E_{4}(q)&\equiv  1+\frac{2}{\zeta(-3)}\sumn \frac{n^{3}q^n}{1-q^n}\\
& \sim~ 1+240 q + 2160 q^2+\cdots 
\end{aligned}
\end{equation}
\noindent
which in the large $N_t$ limit (i.e $q\to 0$) in which we are interested can be
approximated to 1.  Thus we see that the boundary term in the EST action
essentially amounts to an additional correction proportional to $1/R^4$ to the
interquark potential. Looking at eq.(\ref{bound}) we see that $b_2$ is a
dimensional parameter, with dimensions $[\mathrm{length}]^3$.  It is thus
customary to rescale it defining a new dimensionless parameter $\tilde b_2
\equiv \sqrt{\sigma^3}~b_2$

Recent high precision Montecarlo simulations \cite{Brandt:2010bw,
Brandt:2017yzw, Brandt:2018fft, Brandt:2021kvt, Billo:2012da, Bakry:2020ebo,
Bakry:2019cuw} allowed to estimate $\tilde b_2$  for a few LGTs. In particular, for
the $SU(2)$ model in (2+1) dimensions in which we are interested, one finds
$\tilde b_2\sim -0.025$ \cite{Brandt:2010bw, Brandt:2017yzw, Brandt:2018fft,
Brandt:2021kvt}.

We end up in this way with the following asymptotic expression for the interquark potential
\eq
V_{EST}(R)=c+ \sigma R \sqrt{1 -\frac{\pi(D-2)}{12~\sigma R^2}} 
-\tilde b_2\frac{\pi^3}{60 \sqrt{\sigma^3} R^4}
\label{fullV}
\en
with three free parameters: $c$,~$\sigma$ and $\tilde b_2$.

\section{Results}

\subsection{The pure gauge case: analysis of the $\beta_h=0$ data}

As a preliminary step test we first studied the static potential in the
$\beta_h=0$ case in which no string breaking is present and compared our
results with those of \cite{Caselle:2004er} whose simulations were performed at
the same $\beta=9$
value.

We fit the data with eq.(\ref{fullV}) keeping $c$,~$\sigma$ and $\tilde b_2$ as
free parameters in the range $R_{min}\leq R \leq R_{max}$.  We studied
different values of $R_{min}$ in the range $3\leq R_{min} \leq 12$ and
fixed\footnote{We also performed a set of fits varying $R_{max}$ in order to
test for the possible presence of finite size effects. We verified that (with
the exception of the data at $\beta_h=0.95$ which we shall discuss in detail
below) there was no signature of finite size corrections and in the following
we shall only report the results for $R_{max}=20$.} $R_{max}=20$. 
We also performed a set of fits in the same range using the Cornell form of the
potential
\begin{equation}
\label{unb}
V_{\mathrm{Cornell}}(r)=c + \sigma R  -\frac{k_L \pi}{24 R}
\end{equation}
keeping $c$,~$\sigma$ and $k_L$ as free parameters. This expression coincides
with the free bosonic potential eq.(\ref{freeV}) when $k_L=1$ and allows us to
test the improvement of the Nambu-Goto action with respect to the free bosonic
approximation in describing the data. 

\begin{figure}[tbh] 
\includegraphics[width=0.85\columnwidth, clip]{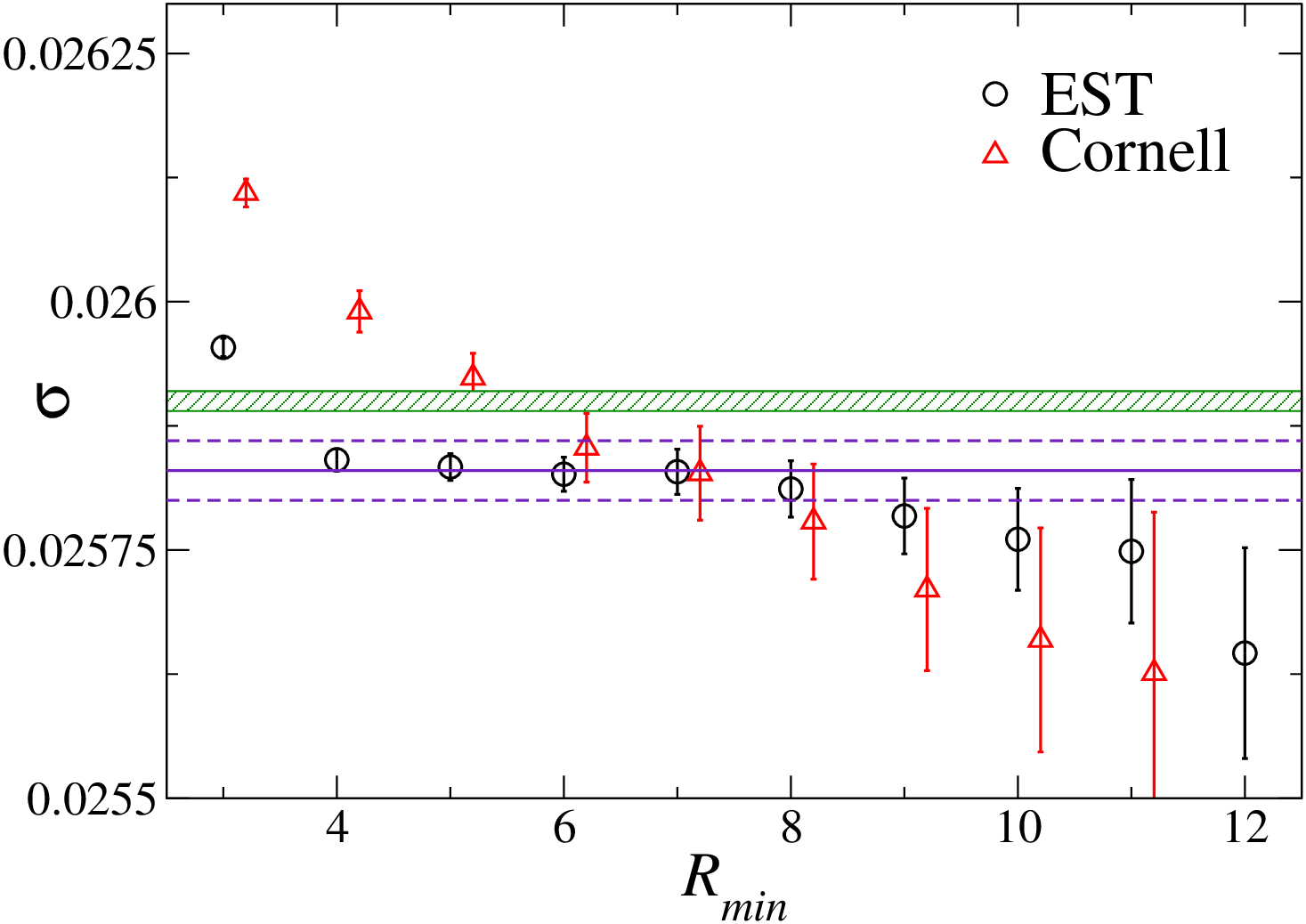}
\includegraphics[width=0.85\columnwidth, clip]{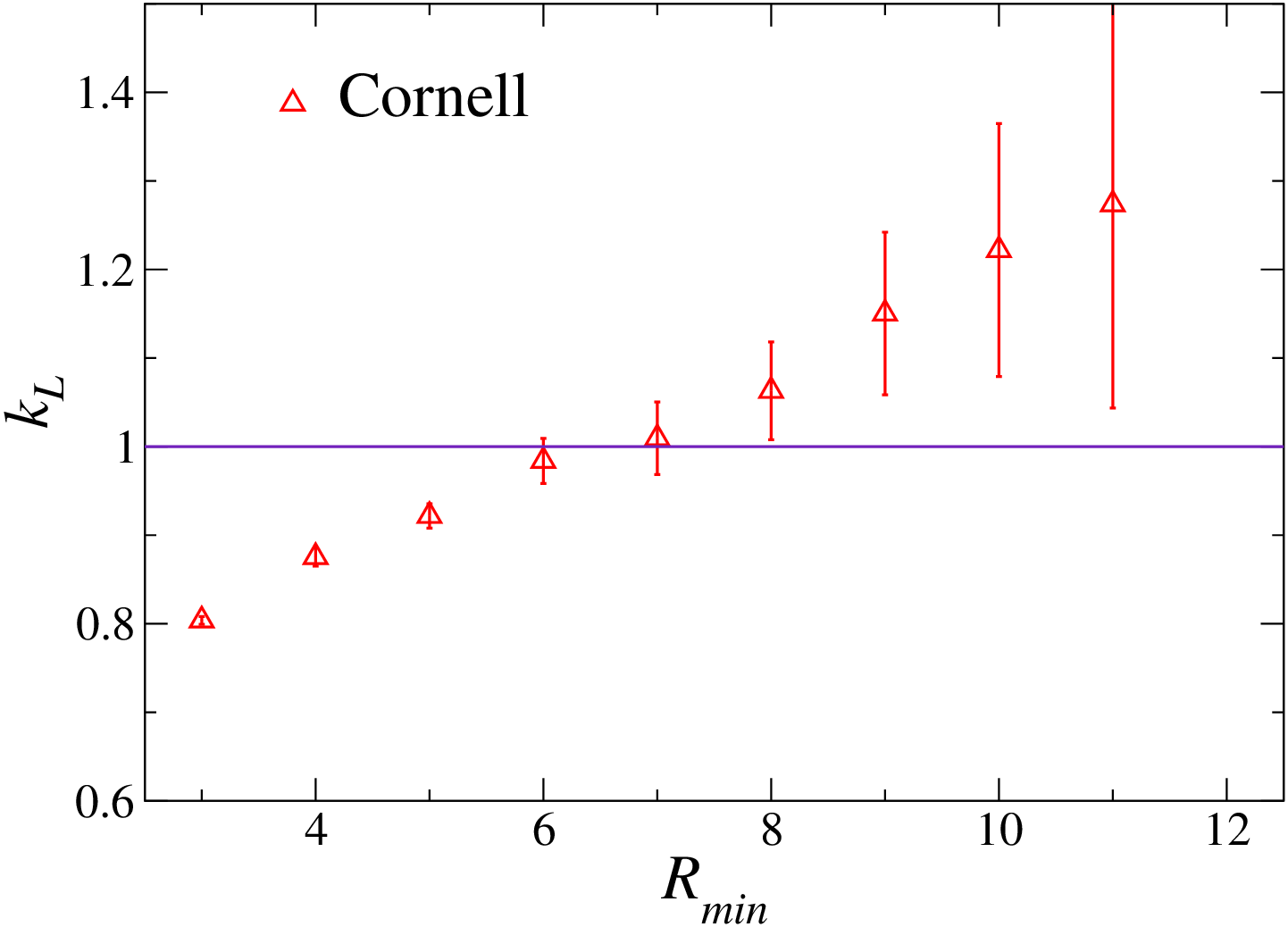}
\caption{Estimates of $\sigma$ (upper panel) and of the Luscher term (lower
panel) for several fitting ranges (data up to $20$ lattice spacings included)
for the case $\beta=9.0$, $\beta_h=0$, $L=42$. As our final estimate 
for the string tension we take
$\sigma=0.02583(3)$, which is also indicated by the horizontal strip.  The
horizontal green band corresponds to the result reported in
\cite{Caselle:2004er}, obtained without taking into account the boundary
contribution to the potential.
}
\label{fig:betah0_sigma_lusch}
\end{figure}

\begin{figure}[tbh] 
\includegraphics[width=0.85\columnwidth, clip]{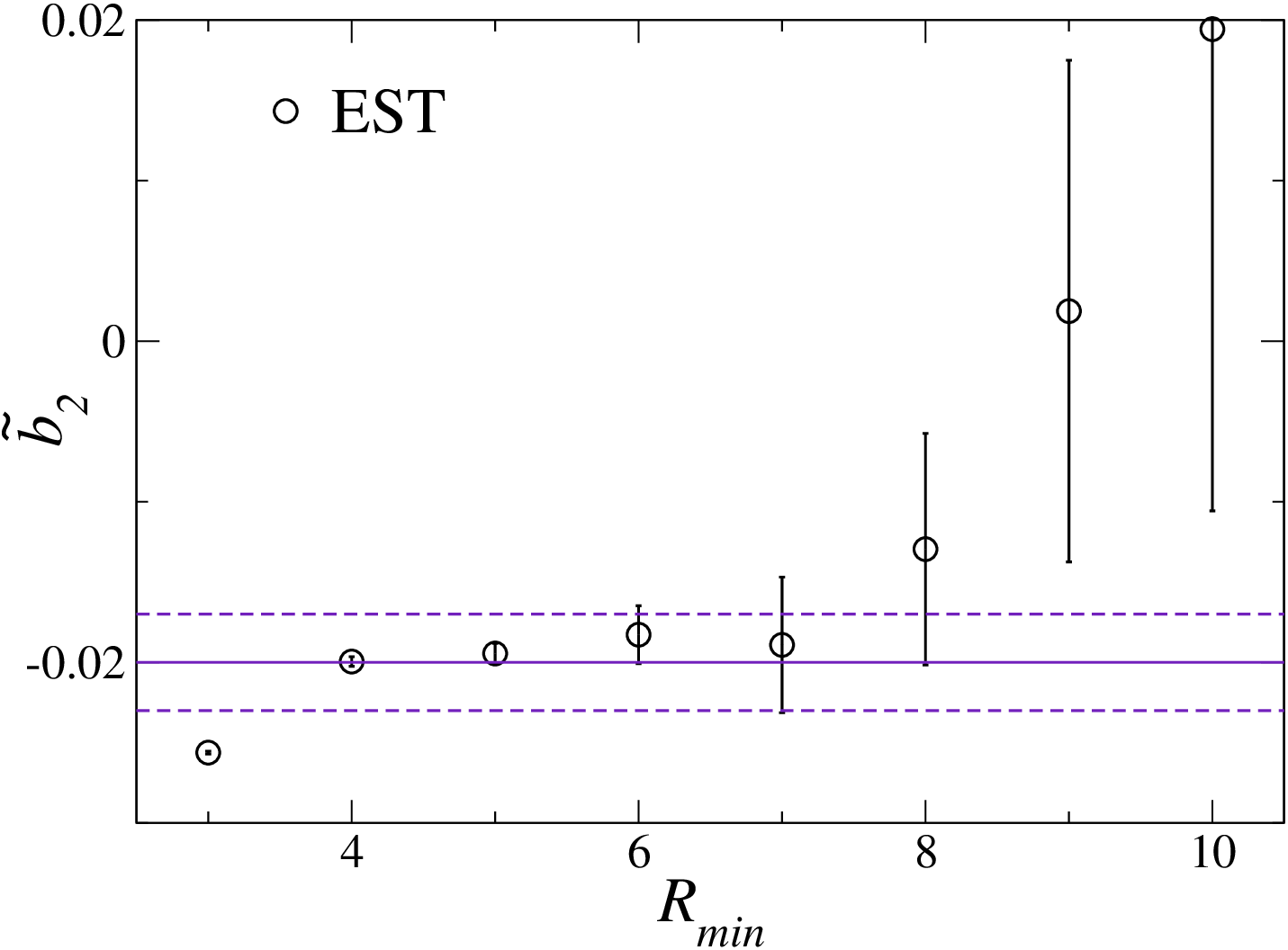}
\caption{Estimates of the coefficient $\tilde b_2$ of the boundary term in
$V_{est}$ for $\beta_h=0$ (data up to $20$ included). We report $\tilde
b_2=-0.020(3)$ as our final estimate.
}
\label{fig:betah0_boundary}
\end{figure}

Results of the fits are reported in fig.s \ref{fig:betah0_sigma_lusch}  and
\ref{fig:betah0_boundary} and, for the EST potential, in table \ref{tab1}.  In
order to give a feeling of the magnitude of the boundary term, we report in the
table the coefficient $\kappa_B$ of the $1/R^4$ correction in the potential
instead of $\tilde b_2$, with:
\begin{equation} \label{kappaB}
\kappa_B\equiv - \tilde b_2\frac{\pi^3}{60 \sqrt{\sigma^3}}
\end{equation}

\begin{table}[tb]
\begin{center}
\begin{tabular}{llllll}
$R_{\mathrm{min}}$ & $\sigma$      & $c$          & $\kappa_B$            & $\chi^2$/dof & dof \\
3                  & 0.0259540(93) & 0.187083(94) & \phantom{-}3.1670(92) & 25.9         & 15  \\
4                  & 0.025841(11)  & 0.18854(10)  & \phantom{-}2.482(30)  & 0.70         & 14  \\
5                  & 0.025834(14)  & 0.18863(16)  & \phantom{-}2.421(76)  & 0.69         & 13  \\
6                  & 0.025826(17)  & 0.18873(22)  & \phantom{-}2.28(22)   & 0.70         & 12  \\
7                  & 0.025829(23)  & 0.18870(31)  & \phantom{-}2.35(52)   & 0.77         & 11  \\
8                  & 0.025811(28)  & 0.18896(40)  & \phantom{-}1.61(90)   & 0.74         & 10  \\
9                  & 0.025784(38)  & 0.18941(58)  & -0.2(1.9)             & 0.69         & 9   \\
10                 & 0.025761(51)  & 0.18981(82)  & -2.4(3.8)             & 0.72         & 8   \\
11                 & 0.025749(72)  & 0.1900(12)   & -3.9(7.3)             & 0.82         & 7   \\
12                 & 0.02565(11)   & 0.1919(16)   & -20(14)           & 0.66         & 6  
\end{tabular}
\end{center}
\caption{Results of the fit of the $\beta_h=0$ data to the $V_{EST}$ potential
for various values of $R_{min}$ and $R_{max}=20$.
$\kappa_B$ is defined in eq.~\eqref{kappaB}.
} 
\label{tab1}
\end{table}

Looking at these data we see a few interesting results:
\begin{itemize}
\item
Inserting in the expression for $R_c$ (see eq.~(\ref{Rc})) the best fit value of
$\sigma$ we find $R_c\sim 3.2$,  thus the range of validity of the EST
potential  is for $R \geq R_{min}=4$. Remarkably enough we find a very good
$\chi^2$ using all the data within the range of validity of EST. Moreover if we
try to add also $R=3$ we have a jump in the $\chi^2$ (which is also associated
with a jump in the value of $\sigma$). Notice  also the remarkable
stability of this result for different choices of $R_ {min}$. As a final result,
taking into account the systematics connected to the choice of the fit range, we
report $\sigma=0.02583(3)$ (shown also in fig.~\ref{fig:betah0_sigma_lusch}).

\item
The value of $\sigma$ that we find is compatible with the scale setting result
$\sigma=0.0262(8)$ of eq.(\ref{scale_setting}). It slightly disagrees with the
estimate of \cite{Caselle:2004er}: $\sigma=0.02590(1)$ which however was
obtained without keeping into account the boundary correction which, as we
shall see below,  is mandatory to correctly fit the data.
\item
 The inclusion of the boundary correction turns out to be  mandatory to fit the
data down to $R_{min}=4$. Any attempt to fit the data without it leads to
unacceptable values of the reduced $\chi^2$ for $R_{min}=4$. The relevance of
the boundary term decreases as we increase $R_{min}$ and becomes negligible
starting from $R_{min}=8$, a result which could have been anticipated by a
direct evaluation of the size of the correction.  Moreover we see, looking at
fig.\ref{fig:betah0_boundary} that in the range $4\leq R_{min}\leq 8$ our
estimate of $\tilde b_2$ is very stable, and keeping into account the
systematic uncertainties of the fit 
we may quote as our final result $\tilde b_2\sim -0.020(3)$ which is fully
compatible with the one obtained by Brandt $\tilde b_2\sim -0.025(5)$
\cite{Brandt:2010bw,Brandt:2017yzw,Brandt:2018fft,Brandt:2021kvt}.
\end{itemize}

It is very interesting to compare these results with those obtained with the
Cornell potential of eq.(\ref{unb}). Looking at fig.s
\ref{fig:betah0_sigma_lusch} we see that the string tension shows a clear trend
to decrease as $R_{min}$ increases and at the same time the coefficient of the
L\"uscher term, which for $R_{min}=4$ is $\approx 10\%$ below the correct result,
increases with $R_{min}$.  The results of the fits stabilize around $R_{min}=8$,
a value for which higher order corrections beyond the gaussian term become
negligible.  We learn from this analysis that for small values of $R$ both the
boundary term and the higher order corrections of the Nambu-Goto action are
mandatory to fit the data, that they have a comparable size and thus must both
be included in the fit.

\subsection{String breaking:  analysis of the $\beta_h\not= 0$ data}
\label{sect:4.2}

Looking at fig.\ref{fig:pot} we see that for the first three values of
$\beta_h$ the shape of the potential is very similar to the unperturbed one.
This suggests trying the same fitting function $V_{EST}$ also for these cases.
In the $\beta_h=0.93$ case this works perfectly, with a reduced $\chi^2=0.85$
in the whole range $4\leq R \leq 20$ and best fit values for the parameters
with a statistical uncertainty similar to that of the unperturbed ones (see the
second line of tab.\ref{tab2}).  However, already for the next value of
$\beta_h$ the same procedure no longer works, meaning that, even if they are
not visible in the plot, the effects of string breaking are already affecting
the large distance behaviour of the potential.  For $\beta_h=0.95$ we can still
control these deviations by cutting the fit to the value $R_{max}=14$ and we
find again values of the parameters similar to the unperturbed ones (see the
third line of tab.\ref{tab2}), but this is no longer possible for the
subsequent values.

\begin{table*}[tbh]
\begin{center}
\begin{tabular}{llllllll}
$\beta_h$ & $\sigma$       & $c$          & $\kappa_B$ & $M$         & $E_{sb}$    & $\chi^2$/dof & dof   \\
0         & 0.025841(11)   & 0.18854(10)  & 2.482(30)  &             &             & 0.70         & 14  \\
0.93      & 0.0254058(56)  & 0.188492(50) & 2.5121(80) &             &             & 0.85         & 14  \\
0.95      & 0.0252119(81)  & 0.188815(71) & 2.515(11)  &             &             & 0.86         & 8  \\
0.97      & 0.024941(15)   & 0.189111(55) & 2.5828(94) & 0.0081(21)  & 0.6836(91)  & 0.75         & 12  \\
1.00      & 0.023842(36)   & 0.19380(19)  & 2.456(59)  & 0.02541(55) & 0.54818(69) & 1.04         & 11  \\
1.02      & 0.02172(27)    & 0.2042(14)   & 1.77(27)   & 0.0312(14)  & 0.44064(70) & 1.34         & 11  
\end{tabular}
\end{center}
\caption{Results of the fit with the $V_{sb}$ potential for various values
of $\beta_h$ and the choices of $R_{min}$ and $R_{max}$ discussed in the text.
}
\label{tab2}
\end{table*}

We may address the string breaking phenomenon, following
\cite{Philipsen:1998de,Knechtli:1998gf,Knechtli:2000df}, assuming a mixture of
a ``string state'' described by $V_{EST}(R)$ and a ``broken string state''
$E_{sb}(R)$ (for a different approach see e.g.
\cite{Antonov:2003ir,Antonov:2005rk}). The simplest way to model this mixture
is by diagonalizing the
$2\times 2$ matrix
\begin{equation}\label{2matrix}
\left( \begin{array}{cc} V_{EST}(R) & M(R) \\ M(R) & E_{sb}(R) \end{array}\right)\ ,
\end{equation}
where $V_{EST}(R)$ is the energy of the ``string state'', $M(R)$ is a mixing
term (a priori dependent on $R$) and $E_{sb}(R)$ is the energy of the ``broken
string state''. We shall assume in the following, as a first approximation,
that both $M$ and $E_{sb}$ have a negligible dependence on $R$ and shall take
them as constants (we shall comment on this approximation at the end of this
section).

The eigenvalues of the mixing matrix are
\begin{equation}
V_{\pm}(R)=\frac{V_{EST}(R)+E_{sb}\pm \sqrt{(V_{EST}(R)-E_{sb})^2+4M^2}}{2}
\end{equation}
with the fundamental state, the one that we observe in our simulations, being
associated to the minus sign.  We thus fitted the Polyakov loop correlators
with the ``string breaking potential'' $V_{sb}(R)$
\begin{equation}
V_{sb}(R)=\frac{V_{EST}(R)+E_{sb}- \sqrt{(V_{EST}(R)-E_{sb})^2+4M^2}}{2}
\end{equation}
with five degrees of freedom: the three parameters contained in $V_{EST}$:
$c$,~$\sigma$ and $\kappa_B$ and the two new parameters $M$ and $E_{sb}$.
Despite the large number of free parameters the fits turned out to be very
stable for all the values of $\beta_h$ that we studied, with good $\chi^2$
values in the whole range $4\leq R \leq 20$ for $\beta_h=0.97$ and in the range
$5\leq R \leq 20$ for $\beta_h=1.00$ and $1.02$. For these values of
$\beta_h$, including in the fit also the point at $R=4$ led to values of the
reduced $\chi^2\sim 2$. 

By exploring the parameter space in the vicinity of the best fit values
reported in tab.\ref{tab2}, we realized that they correspond to very deep and
narrow minima in the parameter space and this probably explains the stability
of the fits. To better asses the reliability of the fitting procedure we also
fixed one of the parameters to its 1$\sigma$-deviation value, checking the
stability of the remaining parameters, whose variations are taken as estimators
of the systematics.  In this way we get the for the three relevant physical
quantities $\sigma$, $\tilde{b}_2$ and $E_{sb}$ the results reported in
tab.\ref{tab3}.  For $\beta \ge 0.97$ the errors quoted in tab.\ref{tab3} for
$\sigma$ and $E_{sb}$ are dominated by the uncertainty on the mixing parameters
$M$. The error on $\tilde b_2$ is instead dominated by the value of $R_{min}$
chosen in the fit and we estimated it combining the values obtained up to
$R_{min}=6$.

Following \cite{Philipsen:1998de} we can extract from the results of the fit a
rough estimate of the ``string breaking threshold'' $R_{sb}$ defined as the value
of $R$ for which the extrapolation of $V_{EST}$ crosses $E_{sb}$.  We report
these values in the last column of tab.\ref{tab3}.  For $\beta_h=1.10$ we could
not perform the analysis since we have too few values of $R$ before the string
breaking threshold.

\begin{table}[tbh]
\begin{center}
\begin{tabular}{lllll}
$\beta_h$  & $\sigma$   & $\tilde{b}_2$   & $E_{sb}$                     & $R_{sb}$ \\
0          & 0.02583(3) & 0.020(3)        &                              &          \\
0.93       & 0.02541(1) & 0.019(2)       &                              &           \\
0.95       & 0.02523(3) & 0.019(3)        &                              &           \\
0.97       & 0.02493(3) & 0.020(1)        &  0.683$^{+0.018}_{-0.009}$   &  $\sim 20$  \\
1.00       & 0.02386(8) & 0.019(3)        & 0.548(2)                     &  $\sim 15$  \\
1.02       & 0.0214(8)  & 0.006(15)       &  0.440(2)                    &  $\sim 10.5$ 
\end{tabular}
\end{center}
\caption{Best fit estimates for $\sigma$, $\tilde{b}$  and $E_{sb}$ for various
values of $\beta_h$. The quoted uncertainties keep into account various
systematic effects in the fits, as discussed in the text. Note that the error
on $E_{sb}$ for $\beta_h=0.97$ is quite asymmetric, likely due to the fact that 
$R_{sb}$ is just on the boundary of the fit range.
}
\label{tab3}
\end{table}

A few comments are in order on these results
\begin{itemize}
\item
As $\beta_h$ increases the string breaking scale $R_{sb}$ decreases and will
eventually become  smaller than the critical radius $R_c$ of EST.  When
$R_{sb}<R_c$ we obviously do not expect an EST regime at short distance.
Looking at our data (see fig.\ref{fig:pot}) this threshold in $\beta_h$ seems
to be reached at $\beta_h=1.10$ for which $R_{sb}\sim 4$. Beyond this value we
may consider the string breaking process to be completed.

\item
It is interesting to see that for $\beta_h=0.97$ (for which $R_{sb}\sim 20$)
looking at the data apparently there seems to be no evidence of string
breaking, (see fig.\ref{fig:pot}), however we have seen from the above analysis
that this impression is wrong and that, without keeping into account the mixing
with the $E_{sb}$ term it would be impossible to fit the data (even if the
string breaking threshold is larger than the set of data included in the fit).

\item
The string tension shows a smooth decreasing trend as $\beta_h$ increases.
This trend is small in magnitude, but definitely larger than the uncerteinties.
It can be used to construct lines of ``constant physics'' in the
$(\beta,\beta_h)$ phase diagram. These lines almost (but not exactly) coincide
with vertical lines ($\beta=const$). The value of the boundary term is almost
constant within the errors along the whole line and agrees with the estimate
obtained by Brandt in the $\beta_h=0$ case.

\item
As we mentioned above for $\beta_h=1.00$ and $1.02$ including in the fit also
the point at $R=4$ led to an increase in the values of the reduced $\chi^2$.
This  small deviation of the  $R=4$ value with respect to the EST prediction
could be the signature of a short distance dependence of $E_{sb}$ on $R$.
Indeed it is conceivable to have in the broken string potential a massive
excitation which would show up in a term of the type
\eq
E_{sb}(R)= E_{sb}+ A e^{-mR}
\en
unfortunately, the range of our data does not allow to extract such a massive
excitation, which however could become visible performing simulations at a
larger value  $\beta$, with a smaller value of the lattice spacing.
\end{itemize}

We conclude this section with a few comments on the $R$-independence of the
entries $M(R)$ and $E_{sb}(R)$ of the mixing matrix in eq.~\eqref{2matrix}.
While it is natural to expect the $R$-dependence of these terms to be weaker
than the one in $V_{EST}(R)$, the fact that our data are perfectly reproduced
by completely neglecting this dependence could seem surprising and, maybe, an
indication that the model studied is somehow pathological. It is thus
reassuring that the same is true also for the case of QCD studied in
Ref.~\cite{Bulava:2019iut}, where however only the linearly rising part of
$V_{EST}(R)$ was used to fit the data. To unambiguously identify the
$R$-dependence of $M(R)$ and $E_{sb}(R)$ much higher accuracy seems to be be
required, likely together with a careful investigation of the excited states.

\section{Concluding remarks}

As we have seen, the formalism of the mixing matrix allows to disentangle in a
clean and precise way the string breaking potential from the confining one and
allows to study fine details of both potentials.  Remarkably enough, the EST
picture seems to describe well the data even in presence of string breaking.
In particular, the boundary term and the higher order terms beyond the gaussian
one in the Nambu-Goto action seem not to be affected by the string breaking and
are clearly visible in the fits.

This is particularly important since, as we have seen, when string breaking
occurs the confining potential can be studied only at short distance, below the
string breaking threshold, and it is exactly in this regime that higher order
terms of the EST like the next to gaussian terms of the Nambu-Goto action and
the boundary term, become particularly important and cannot be neglected.  The
fits performed with the gaussian term only (see
fig.\ref{fig:betah0_sigma_lusch}), show that neglecting these corrections would
lead to a wrong estimate of the string tension (and of the Luscher term
itself).  We think that this is an important lesson to keep in mind when
looking at the interquark potential in QCD for which the string breaking scale
is of the same size of the intermediate values of $\beta_h$ that we studied in
this paper. Since we are by now entering the precision era of lattice
simulations for phenomenology
\cite{Bicudo:2019ymo,Bicudo:2020qhp,Bulava:2019iut}, this type of corrections
will become more and more important and should be kept into account to reach
the correct phenomenological estimates of the interquark potential.

\emph{Acknowledgments} Numerical simulations have been performed on
the CSN4 cluster of the Scientific Computing Center at INFN-PISA.


%

\end{document}